\documentstyle[aps,preprint]{revtex}

\begin{document}
\draft
\preprint{\begin{tabular}{l}
Imperial/TP/97-98/45\\
LPTENS-98/21 \\
ULB-TH-98/09
\end{tabular}}

\title{ Anti-de Sitter/CFT Correspondence in Three-Dimensional
  Supergravity} 
  
\author{M\'aximo Ba\~nados\,$^{a,b}$, Karin
  Bautier\,$^{c}$, Olivier Coussaert\,$^{c}$, Marc
  Henneaux\,$^{b,c,d}$\\ and Miguel Ortiz\,$^{e}$}

\address{$^a$Departamento de F\'{\i}sica Te\'orica, Universidad de
  Zaragoza,
  Ciudad Universitaria 50009, Zaragoza, Spain,\\
  $^b$Centro de Estudios Cient\'\i ficos de Santiago,
  Casilla 16443, Santiago 9, Chile.\\
  $^c$Physique Th\'eorique et Math\'ematique, Universit\'e Libre de
  Bruxelles,
  Campus Plaine C.P. 231, B--1050 Bruxelles, Belgium,\\
  $^d$Laboratoire de Physique Th\'eorique de l'Ecole Normale 
Sup\'erieure\footnote{Unit\'e Propre du Centre National de la
Recherche
Scientifique, associ\'ee \`a l'Ecole Normale Sup\'erieure
et \`a l'Universit\'e de Paris-Sud},
  24, rue Lhomond, 75230 Paris Cedex 05, France, \\
  $^e$Theoretical Physics Group, Blackett Laboratory, Imperial
  College, London SW7 2BZ, UK.}
 \maketitle 
 
 \newpage
 
\begin{abstract}
  Anti-de Sitter supergravity models are considered in three
  dimensions.  Precise asymptotic conditions involving a chiral
  projection are given on the Rarita-Schwinger fields.  Together with
  the known boundary conditions on the bosonic fields, these ensure
  that the asymptotic symmetry algebra is the superconformal algebra.
  The classical central charge is computed and found to be equal to
  the one of pure gravity.  It is also indicated that the asymptotic
  degrees of freedom are described by $2D$ ``induced supergravity" and
  that the boundary conditions ``transmute" the non-vanishing
  components of the WZW supercurrent into the supercharges.
\end{abstract}

\newpage

%\vskip0.5cm

%\pacs{04.65.+e;11.30.ly;04.20.Fy}

%\narrowtext

\section{Introduction}

It has been pointed out in \cite{Brown-Henneaux} that the asymptotic
symmetry group of anti-de Sitter gravity in three dimensions is the
conformal group in two dimensions with a central charge $c=3l/2G$. The
emergence of the conformal group at infinity can be understood either
in terms of Penrose conformal treatment of infinity \cite{Penrose} or
by working out explicitly the boundary conditions and solving the
asymptotic Killing equations \cite{Brown-Henneaux}.  It is a purely
asymptotic phenomenon, in the sense that the infinite-dimensional
conformal group in two dimensions is not the isometry group of any
$3D$ background geometry.  This is one feature that makes the
three-dimensional case particularly interesting and which actually
allows for a non trivial central charge in the dynamical realization
of the asymptotic symmetry algebra \cite{Brown-Henneaux}.

Another interesting feature of three-dimensional gravitational
theories is that they have no bulk degrees of freedom, so that the
r\^ole of the boundary degrees of freedom in the adS/CFT
correspondence \cite{W1,MS,Maldacena,GKP,W2} can be investigated more
easily.  That the boundary degrees of freedom may be quite significant
has been stressed recently in \cite{MV}.  The pure gravitational case
has been analysed in \cite{CHvD}, where it was shown that the boundary
dynamics at infinity is described by ``induced $2D$ gravity"
(Liouville theory) up to terms involving the zero modes and the
holonomies that were not worked out.

The purpose of this paper is to extend the analysis of
\cite{Brown-Henneaux,CHvD} to the supersymmetric
context, which is known to play a central r\^ole in black hole
physics.  The new non-trivial ingredient to be fed in is the precise
asymptotic behavior of the Rarita-Schwinger fields, which must be
compatible with the symmetries.  In particular, one must understand
how the boundary conditions implement two-dimensional supersymmetry at
infinity.

The supersymmetry properties of $3D$ black holes were investigated in
\cite{CH}, assuming the existence of asymptotic conditions on the
Rarita-Schwinger fields fulfilling the required properties. However,
the asymptotic conditions in question were not given.  The
main object of our paper is to fill this gap, which appears necessary
since otherwise, the discussion of the asymptotic dynamics remains
rather formal.  We also verify that the given fall-off conditions
reduce the theory to induced $2D$ supergravity.  The symmetry algebra
is the super-conformal algebra with unchanged central charge
$c=3l/2G$.  The boundary conditions involve a chiral projection, in a
way very similar to what has been discussed for Dirac fields in
\cite{HS}.

\section{Boundary conditions}

AdS supergravity in three dimensional spacetime can be written as a
Chern-Simons theory \cite{Achucarro-Townsend}. We shall adopt the
Chern-Simons point of view from the outset and consider almost
exclusively the (1,1) theory.  The relevant group is $OSp(1|2)\times
OSp(1|2)$.  The $(1,1)$-supergravity action is
\begin{equation}
I[A,\psi;\tilde{A}, \tilde{\psi}] = I[A,\psi] - I[\tilde{A},
\tilde{\psi}]
\label{a1}
\end{equation}
where $I[A, \psi]$ and $I[\tilde{A}, \tilde{\psi}]$
are the Chern-Simons actions for the
supergroup $OSp(1|2)$,
\begin{equation}
I[A,\psi] = \frac{k}{4\pi} \int \left[ \mbox{Tr}(A dA + \frac{2}{3}
A^3)
+ i \bar \psi \wedge D \psi \right]
\label{a2}
\end{equation}
(with a similar expression for $I[\tilde{A}, \tilde{\psi}]$).  Here $A
= (1/2) A^a \gamma_a$ and the symbol Tr stands for the trace in the
spinorial representation of $SO(2,1)$ generated by $J_a = (1/2)
\gamma_a$ (our conventions are summarized in \cite{con}).  The
constant $k$ is related to the $3D$ Newton constant $G$ and the
anti-de Sitter radius $l$ through $k = l/4G$.

Assuming that the topology of the three dimensional manifold $M$ is
$\Sigma \times \Re$, the action (\ref{a2}) can be recast in
Hamiltonian form as
\begin{equation}
I =  \int \left[ -\frac{k \epsilon^{ij}}{4\pi} \left(
\frac{\eta_{ab}}{2}  A^a_i \dot A^b_j + i \bar\psi_i \dot \psi_j
\right) -  A_0^a {\cal G}_a - \bar \psi_0  {\cal S} \right]
\label{a3}
\end{equation}
where the constraints ${\cal G}_a$ and ${\cal S}$ are given by
%\begin{equation}
${\cal G}_a \equiv -(k \epsilon^{ij}/8\pi) \eta_{ab} (F^b_{ij} -i
\bar \psi_i \gamma^b \psi_j ) = 0 $, %\approx
${\cal S} \equiv -(ik/2\pi)\epsilon^{ij} D_i \psi_j = 0 $ 
%\end{equation}
and satisfy the $OSp(1|2)$ algebra in the Poisson brackets
$[A^a_i,A^b_j] = (4\pi/k) \eta^{ab} \epsilon_{ij}$,
$\{\psi^\alpha_i,\psi^\beta_j \} = (2\pi i/k) \epsilon_{ij}
(\gamma_0)^{\alpha\beta}$ which are derived from (\ref{a3}). 
The canonical generator of the gauge transformations (including the
fermionic ones) is $G(\lambda^a) + S(\rho)$ with
\begin{equation}
G(\lambda^a)   = \int_\Sigma \lambda^a {\cal G}_a + B , \; 
S(\rho) = \int_\Sigma \bar \rho {\cal S} + F \label{S}
\end{equation}
The boundary terms $B$ and $F$ must be chosen so that the generators
$G$ and $S$ have well-defined functional derivatives \cite{RT}, and
their precise form depends on the boundary conditions.

The boundary conditions at infinity on the bosonic fields have been 
given in \cite{Brown-Henneaux} in the metric representation and they 
were
reexpressed in the connection representation in \cite{CHvD}. (See 
\cite{Max,MM} for a different approach to the problem of boundary 
conditions in the connection representation.) The bosonic boundary 
conditions must be supplemented by appropriate boundary conditions on 
the fermionic fields.  We consider only one $OSp(1|2)$ copy, the other 
copy being treated similarly. 

The searched-for boundary conditions can be determined by following
the procedure of \cite{HT}: one starts with the known physical metrics
that should be included in the theory - here, the black hole solutions
\cite{BTZ} - and acts on them with the anti-de Sitter supergroup.
This suggests to adopt for the Rarita-Schwinger fields the following
boundary conditions (in the standard orthonormal frames)
\begin{eqnarray}
\psi_t &\sim& r^{-\frac{1}{2}} [1 + \gamma_1] \chi(t, \phi)
, \label{b1}\\
\psi_\phi &\sim& r^{-\frac{1}{2}} [1 + \gamma_1] \chi(t, \phi)
 , \label{bc}\\
\psi_r &\sim& r^{-\frac{5}{2}} [1 - \gamma_1] \chi_r(t, \phi).
\end{eqnarray}
Apart from an irrelevant replacement of $\gamma_1$ by $ - \gamma_1$
(due to conventions), the boundary conditions differ from those of
\cite{HT} (Eqn. (V.I)) in two respects. First, they involve a slower
rate of decrease at infinity (one less power of $r$).  This was to be
expected since we are one dimension lower and also holds for the
bosonic fields \cite{Brown-Henneaux}.  Second, they have the same
leading order for both $\psi_t$ and $\psi_\phi$.  The equality of the
leading orders of $\psi_t$ and $\psi_\phi$ is consistent with the fact
that the adS Killing spinors of $(1,0)$ supergravity depend only on
$t+ \phi$ \cite{CH}.  The boundary conditions are otherwise identical
and in particular, they crucially involve a projection onto the
eigenspaces of the radial $\gamma$-matrix, which makes the induced
spinors chiral in two dimensions (recall that $\gamma_1$ appears as
the ``$\gamma_5$"-matrix on the surface at infinity).  A similar
phenomenon is described in \cite{HS}.

\section{Asymptotic symmetries}

The algebra of coordinate and supersymmetry transformations that
preserves the boundary conditions (``asymptotic symmetry algebra") is
the infinite-dimensional super-Virasoro algebra.  To discuss this
issue, it is most convenient to work in the superconnection
representation where the transformations take a simpler form.
 
Combining the above boundary conditions on the fermions with those of
\cite{Brown-Henneaux,CHvD} on the bosons, one finds that the
superconnection must satisfy
\begin{eqnarray}
A_v = 0, \ & \ \psi_v = 0, \label{chi} \\
A_r = b^{-1} \partial_r b, \ & \ \psi_r = 0 \label{r} 
\end{eqnarray}
and
\begin{equation}
l A_u =  b^{-1} \left(\begin{array}{cc}
                     0  &  L/k \\
                     1  &   0
                   \end{array} \right) b,
\ \ l  \psi_u =
    b^{-1}\left(\begin{array}{c}
                             Q/k  \\
                                 0
                   \end{array} \right)
\label{A+}
\end{equation}
asymptotically.  Here, $u = t +l \phi$, $v = t-l \phi$ and
$L=L(t,\varphi)$ and $Q=Q(t,\varphi)$ are arbitrary functions 
which will be shown to be equal to the generators of the
super-Virasoro
algebra.  The group element $b(r)$ is equal to
\begin{equation}
b(r) = \left(\begin{array}{cc}  (r/l)^{1/2} &  0 \\
                                            0  &  (r/l)^{-1/2}
                   \end{array} \right)
\end{equation}
and satisfies $b\gamma_0 b = \gamma_0$.  Note that $A_\phi = l A_u$,
$\psi_\phi = l \psi_u$ since $A_v=0$, $\psi_v = 0$.  The other
$OSp(1|2)$ field satisfies analogous boundary conditions, with $u$ and
$v$ interchanged, and depends on two additional functions $\tilde L$
and $\tilde Q$. For positive values of $L_0$ and $\tilde{L}_0$ the
boundary
conditions (\ref{A+}) represent a black hole. The black hole ground
state ($M=0$) is obtained for $L=\tilde L=0$. Anti de Sitter space
corresponds to $L/k=\tilde L/k=-1/4$ and is the only configuration for
which the
holonomies are trivial. (Note, however, that because $A$ is written in
the spinorial representation, the holonomy (in polar
coordinates) is only trivial under a
$4\pi$ rotation).

The fact that the second component of $\psi_u$ is zero just expresses
the chirality condition on the fermion enforced by the boundary
conditions (\ref{b1}) and (\ref{bc}).  One may rewrite the asymptotic
form (\ref{A+}) of the superconnection in terms of supermatrices as
\begin{equation}
\Gamma_u =  b^{-1} \left(\begin{array}{ccc}
                0 &  L/k & Q/(\sqrt{2}k) \\
                1 &  0   & 0   \\
                0 & Q/(\sqrt{2}k) & 0
                \end{array} \right)  b
\label{psi+}
\end{equation}
where $b$ is now the $3 \times 3$ supermatrix obtained by completing
the above $b$ by adding $0$ in the fermionic positions and $1$ in
position $(3,3)$.  The advantage of the connection representation is
that one can completely eliminate the $r$-dependence through the gauge
transformation generated by $b$.  After this gauge transformation has
been performed, all the asymptotically relevant components of the
fields occur at the same order $O(1)$.  Furthermore, because $r$ has
dropped out, the analysis could be carried out in the same way at any
finite value of $r$. A Virasoro algebra for all values of $r$ has been 
investigated in \cite{MM}. 

The most general supergauge transformations that preserve the boundary
conditions (\ref{chi}), (\ref{r}) and (\ref{A+}) are characterized by 
gauge parameters $(\lambda^a, \rho)$
that must fulfill, to leading order, 
\begin{equation}
\lambda(u, r) = b^{-1} \eta(u) b, \ \ \rho(r,u) = b^{-1}
\varepsilon(u)
\label{geneform}
\end{equation}
with
\begin{eqnarray}
\eta^+ &=& \frac{\eta^- L}{k}
 - (1/2) (\eta^-)''+ \frac{iQ\epsilon}{2 k}, \label{p1}  \\ 
\eta^1 &=& - (\eta^-)'      \label{p2}        \\
\varepsilon &=& \left(\begin{array}{c}
           -\epsilon' +\eta^- Q/k \\
           \epsilon
                   \end{array} \right)
\label{res-susy}
\end{eqnarray}
where $'$ denotes derivative with respect to the argument.  We have
expanded the algebra element $\eta$ in the Cartan basis $\eta = \eta^1
J_1 + \eta^+ J_+ + \eta^- J_-$. Equations (\ref{p1})-(\ref{res-susy})
imply that the full residual symmetry can be
expressed in terms of two functions of the lightlike coordinate $u$,
one bosonic ($\eta^-$) and one fermionic ($\epsilon$).  

Any three-dimensional gauge transformation whose parameters 
fulfill  (\ref{p1})-(\ref{res-susy}) asymptotically is called ``an
asymptotic 
symmetry". Two gauge transformations that tend to the same $\eta^-$
and 
$\epsilon$ at infinity should be identified because
they differ by a ``proper gauge transformation",  which it is
legitimate 
to quotient out \cite{RT,Benguria} (as will be clear below, these 
transformations have in particular the same global charges). 
The resulting quotient superalgebra is the
``asymptotic symmetry superalgebra".  

If one computes the graded commutator of two Chern-Simons gauge
transformations fulfilling the above asymptotic conditions and
characterized by asymptotic parameters $(\eta^-_{1}, \epsilon_{1})$
and $(\eta^-_{2}, \epsilon_{2})$,
one finds another such transformation with asymptotic
gauge parameters related to $(\eta^-_{1}, \epsilon_{1})$
and $(\eta^-_{2}, \epsilon_{2})$ {\em exactly according
to the graded commutation rules of the super-Virasoro algebra}.
Hence, after the quotient by the ideal of the
proper gauge transformations (with $\eta^-=0, \epsilon=0$) is
taken, one is left with the super-Virasoro algebra as 
asymptotic symmetry superalgebra.  This
infinite-dimensional algebra
contains $OSp(1|2)$ as a subalgebra when the fermions are
anti-periodic (Fourier modes
$0$ and $\pm 1$ of $\eta^-$ and modes $\pm \frac{1}{2}$ of
$\epsilon$).
Note, in particular, that the Lie algebra commutator $[\lambda^a_{1},
\lambda^b_{2}]$
of two bosonic gauge transformations restricted by
(\ref{geneform}) and (\ref{p1}) reduces at infinity
to the Lie bracket of the residual functions $\eta^-_{1}$ and
$\eta^-_{2}$ viewed as vector fields on the circle.
A similar statement holds for the fermionic sector.

\section{adS central charge}

We now turn to the discussion of the canonical realization of 
the asymptotic symmetry algebra.  As is known, the bracket of 
the canonical generators of the asymptotic symmetries provides a 
projective representation of the algebra 
\cite{Brown-Henneaux,Brown-Henneaux2}.  
To determine the central charges, one must first work out the 
complete form of the generators  (\ref{S}).  This is now possible 
since the asymptotic form of both the fields and the 
symmetry transformations has been obtained. One finds that with the
above 
asymptotic conditions, the boundary terms in the variation of 
the generators (\ref{S}) cancel out if one takes
\begin{equation}
B = \frac{1}{2\pi}\int_{\partial\Sigma} \eta^- L, 
\ \ \ \ F = \frac{-i}{2\pi}\int_{\partial\Sigma}
 \epsilon Q
\label{surfT}
\end{equation}
i.e., the surface terms are precisely $L$ and $Q$ (up to numerical
factors).  We have adjusted the constants in the charges so that these
vanish for the zero mass black hole, which has $L=0$.  The surface
terms (\ref{surfT}) are of course equal to the surface terms
that one would obtain through a more orthodox ``non-Chern-Simons-based"
approach (see \cite{HT} for the four-dimensional treatment).
In particular, the bosonic piece $B$ is equal to the
charge (4.11) of \cite{Brown-Henneaux} written in terms of the metric, while
the fermionic surface term may be re-expressed as
\begin{equation}
 F = \frac{ik}{2\pi}\int_{\partial\Sigma} \bar{\rho} \psi_\phi.
 \end{equation}

Because the components $L$ and $Q$ of the connection that
remain at infinity enter the surface
terms in the canonical generators of the asymptotic symmetries, it is
useful to know how they transform under the asymptotic symmetry group.
The transformation law for the superconnection $\delta \Gamma = D
\Lambda$ yields 
\begin{eqnarray}
\delta L &=& (\eta^- L)' + (\eta^-)' L  - \frac{k}{2} (\eta^-)'''
+(\frac{iQ\epsilon}{2})' + i Q \epsilon'
\label{dL} \\
\delta Q &=&  -k\epsilon'' + L \epsilon
+(\eta^- Q)' + \frac{1}{2} (\eta^-)' Q.
\label{dQ}
\end{eqnarray}
The equations (\ref{dL}) and (\ref{dQ}) indicate that $L$ and $Q$ form
a super-Virasoro algebra.  More importantly, the transformation laws
(\ref{dL}) and (\ref{dQ}) give also the central charge $c$, equal to
$6k$ ($c/12 = k/2$).
 
By using the general argument of 
\cite{Brown-Henneaux2,Brown-Henneaux}, or by
direct calculation, one finds
that the Poisson brackets of the improved generators
(\ref{S}) are
\begin{eqnarray}
~[G(\lambda_{1}),G(\lambda_{2}) ] &=& G([\lambda_{1},\lambda_{2}]) 
- \frac{k}{4\pi}\int_{\partial\Sigma} 
 \eta^-_{1} (\eta^-_{2})'''  \\
~[G(\lambda),S(\rho)] &=& S( \lambda^a \gamma_a \rho)  \\
~\{S(\rho_{1}),S(\rho_{2}) \} &=& G( -i \bar\rho_{1}\gamma^a \rho_{2})
+
\frac{i k}{2\pi}\int_{\partial\Sigma} \epsilon_{1} 
(\epsilon_{2})'' \label{cons-al}
\end{eqnarray}
The central term is just that of (\ref{dL}), (\ref{dQ}).
The Chern-Simons formulation of adS supergravity provides 
a particularly efficient derivation of the adS central charge.

The above algebra involves both the proper gauge symmetries and the
improper ones \cite{RT,Benguria}. The proper gauge symmetries have
(weakly) vanishing Poisson brackets with all the other generators,
i.e.,
form also an ideal in the Poisson sense.  It follows that the
generators of the asymptotic symmetries are ``first class" and
well defined in the reduced phase space obtained by quotientizing
the proper gauge symmetries.  Using standard terminology, they 
are ``observables".  The Poisson bracket (in the reduced phase
space) of these observables is the same as their Poisson
bracket in the original phase space (see e.g. \cite{HT2}).  
That the global charges at the boundary define observables
has been particularly emphasized recently in
\cite{Balachandran}.
The super-Virasoro algebra is therefore
realized in the space of physical observables, where it is
generated by $L$ and $Q$ (the constraints are
zero in the reduced phase space).
After Fourier transformation, the asymptotic superalgebra 
takes the familiar form
(using quantum-mechanical notation and rescaling $Q$ by $\sqrt{2}$),
\begin{eqnarray}
~[L_m,L_n] &=& (n-m) L_{n+m} + \frac{k}{2} n^3 \delta_{n+m,0} 
\label{vis1}\\
~[L_m,Q_n] &=& \left(\frac{1}{2}m - n\right) Q_{m+n} \\
~\{Q_m,Q_n\}&=& 2L_{m+n} + 2k  m^2 \delta_{m+n, 0}
\label{vis3}
\end{eqnarray}
with a central charge equal to $c=6k$. A practical way to factor out 
the proper gauge symmetries is
to fix the gauge in the bulk and use Dirac brackets
\cite{Benguria}. In that case, the reduced brackets in the
above algebra would appear as Dirac brackets.

Note that if we had imposed only the chirality boundary condition
(\ref{chi}), as is usually done in the Chern-Simons $\rightarrow$
chiral WZW reduction, we would have obtained a current
(Kac-Moody) algebra rather than the super-Virasoro algebra. The key
point
leading to the super-Virasoro algebra is the presence of the extra
boundary conditions (\ref{A+}) which ``transmute" the residual gauge
field components functions $L$ and $Q$ into super-Virasoro charges
\cite{P5,P1,Al,Ber,oR}.  These extra boundary conditions express (with
the other boundary conditions given above)
adS asymptotics.
  
{}From the point of view of the chiral WZW theory, this transmutation
can be seen explicitly as follows. Let affine $SL(2,\Re)_k$ be
generated by $J^\pm,J^1$. Impose $J^-=1$ and $J^1=0$ (see
(\ref{A+})). These are second class constraints because their
Poisson bracket is an invertible matrix.  It follows that $J^+=L$
satisfies, in the Dirac bracket, the Virasoro algebra with $c=6k$.
This argument is extended directly to the supergravity theory and will
be given a dynamical
interpretation in the next section.

In (\ref{vis1})-(\ref{vis3}), the fermions can be periodic (index on
$Q$
integer-moded) or anti-periodic (index on $Q$ half-integer moded).
The form of the algebra is adapted to the periodic (``Ramond") case,
which has the zero mass black hole as the $L_0=0$ ground state
\cite{CH}.  The central charge vanishes for the sub-algebra generated
by $(L_0, Q_0)$, which corresponds to the true symmetries of this
background. The anti-de Sitter background has $M= -1$, i.e. $L_0 =
-c/24$.  It is the ground state of the anti-periodic
(``Neveu-Schwarz") sector \cite{CH}.  If one shifts $L_0$ by $c/24$ so
that $L_0$ vanishes on the anti-de Sitter solution, one finds that the
central charge vanishes for $(L_{\pm1}, L_0, Q_{\pm \frac{1}{2}}$),
which are true ($OSp(1|2)$) symmetries of the anti-de Sitter
background).

What we have done for one $OSp(1|2)$ factor can be repeated for the
other $OSp(1|2)$ factor.  The corresponding spinor fields are
projected on the other chirality and one finds another copy of the
super-Virasoro algebra, this time depending on $v = t-l \phi$, with
same
central charge.  The two super-Virasoro algebras give the conformal
superalgebra.

\section{Dynamics at infinity}

As anticipated above,
the emergence of the super-conformal algebra at infinity 
with a non-vanishing central charge can be 
understood at the dynamical level,
in the light of Polyakov's discovery of the ``hidden $SL(2,\Re)$ 
symmetry" of $2D$ gravity \cite{P5,P1,Al,Ber,oR}. The argument runs as
follows 
\cite{CHvD}.  As shown by \cite{W1,MS}, the 
Chern-Simons theory under the boundary condition (\ref{chi}) induces 
the chiral Wess-Zumino-Witten model at the boundary. The corresponding 
Kac-Moody currents are just the $\phi$-components of the connection. 
Combining the two chiral WZW models of opposite chiralities obtained
from each $SL(2,\Re)$-factor, one finds a non-chiral $SL(2,R)$ WZW
theory 
(modulo zero modes and holonomies not discussed here because
they affect neither the asymptotic symmetry nor the central charge).
The constraints on the Kac-Moody currents arising from
the anti-de Sitter asymptotics
lead then to $2D$-gravity \cite{CHvD}.

In a similar way, the further constraint that the component of the
Kac-Moody current along the fermionic generator $f$ vanishes (see
(\ref{A+})) turns out to be precisely the constraint that reduces the
WZW theory based on the supergroup $OSp(1|2)$ to chiral $2D$
supergravity \cite{P1,BersOo2,susy,Inami}.  Although the $OSp(1|2)$-WZW theory
is not superconformal, the resulting theory is.  What happens is that
the other component (along $e$) of the fermionic Kac-Moody
supercurrent is ``transmuted" into the super-Virasoro generator since
its transformation law becomes (\ref{dQ}) once the gauge parameters
are restricted by (\ref{p1}) and (\ref{res-susy}).  From the WZW point
of view, supersymmetry on the worldsheet arises therefore in a non
trivial way. It is rather interesting that these features are in fact
all contained in the $3D$ boundary conditions expressing anti-de
Sitter asymptotics, thus the adS/CFT correspondence is explicit in
this context.  Bringing in the other $OSp(1|2)$ factor leads to the
non-chiral $(1,1)$-supergravity, which is described, in the $2D$
super-conformal gauge, by super-Liouville theory.  We have explicitly
checked, using the Gauss decomposition for $OSp(1|2)$ and following
the same lines as in \cite{CHvD}, that the $3D$ supergravity
action(\ref{a1}) yields the super-Liouville action (up to zero modes
and holonomies that we have not explicitly worked out).

\section{Conclusions}

We have shown in this paper that the anti-de Sitter 
boundary conditions in $(1,1)$
$3D$-supergravity theory lead to an asymptotic symmetry algebra which
is (twice) the super-Virasoro algebra with a central charge equal to
$6k$.  The precise boundary conditions given here on the spinors
involve
a chiral projection and legitimate the assumptions of \cite{CH}.

The appearence of the Virasoro algebra as the boundary symmetry
algebra of anti-de Sitter space is purely kinematical in the sense
that the only ingredients that enter the derivation of both the
symmetry algebra and the central charge in (\ref{vis1}) are (i) the
asymptotic boundary conditions that dictate the approach to anti-de
Sitter; and (ii) the fact that the surface terms at infinity in the
Virasoro generators involve only the (bosonic) gravitational
variables, i.e., the triads and the spin connection.  Any theory with
these features will have the same central charge in the commutator
involving two $L_n$'s.  In particular the extended
$(p,q)$-supergravity models fulfill these properties and have as
asymptotic symmetry algebra the relevant graded extension of the
conformal group with Virasoro algebra fulfilling (\ref{vis1}) with
same $c=3l/2G$.   The triads, spin connection and spinor
fields obey the same asymptotic conditions as above, while
the $SO(N)$-connection $A^{ij}_\mu$ fulfills $A^{ij}_v = 0 = A^{ij}_r$,
$A^{ij}_u = T^{ij}(t,\phi)$.
Of course, since the generators in
these extended superconformal algebras contain, besides the Virasoro
generators, only the $N$ supercharges of conformal
spin $3/2$ and the $SO(N)$-currents of conformal spin
$1$, with no generator of lower conformal spin,
the supersymmetric extensions in question are those
described in \cite{Kni,Bersh2,Fradkin}.  These algebra
close quadratically 
in the $SO(N)$-currents, except for $N=2$ and $N=4$
(with boundary conditions breaking $SO(4)$ to one of its $SU(2)$ 
subgroups), for which one recovers the linear algebras
of \cite{Adem}.  The Hamiltonian reduction of the
corresponding WZW models has been analyzed in \cite{Ito,Sevrin}.

As observed in \cite{Strominger} the degeneracy of states for a 
conformal field theory with this central charge gives rise to, under 
appropriate conditions, exactly the Bekenstein-Hawking entropy for the 
$2+1$ black hole (see also \cite{Birmingham}). 
An earlier statistical description of the $2+1$ black hole 
entropy was given by Carlip \cite{Carlip} in terms of horizon degrees 
of freedom. For further work in these directions, see 
\cite{MM,Bala,Kal,MSt,Mar,IPZ,BBG,Cvetic}.  In view of the relevance of the
$2+1$ black hole to higher dimensional ones \cite{U1,U2}, this
question clearly deserves further study.

\section*{Acknowledgements} 
Useful discussions with Bernard Julia
and Steve Carlip are gratefully acknowledged.
M.H. thanks 
the Laboratoire de Physique Th\'eorique de l'Ecole Normale
Sup\'erieure for kind hospitality while this work was
completed. K. B. is ``Chercheur F.R.I.A." (Belgium).
%The ``Laboratoire de Physique Th\'eorique de l'Ecole Normale
%Sup\'erieure" is a ``Unit\'e Propre du Centre National de la Recherche
%Scientifique, associ\'ee \`a l'Ecole Normale Sup\'erieure
%et \`a l'Universit\'e de Paris-Sud".

%-----------------------------------------


\begin{references}

\bibitem{Brown-Henneaux} J.D. Brown and M. Henneaux, {\em Commun.
Math. Phys.} {\bf 104}, 207 (1986).
\bibitem{Penrose} R. Penrose, in {\em Relativity, Groups and
Topology},
eds C. De Witt and B. De Witt, Gordon and Breach (New York: 1964).
\bibitem{W1} E. Witten, {\em Commun. Math. Phys.} {\bf 121}
(1989) 351. 
\bibitem{MS} G. Moore and N. Seiberg, {\em Phys. Lett.} {\bf 220 B}
(1989)422; S. Elitzur, G. Moore, A. Schwimmer and N. Seiberg
{\em Nucl. Phys.} {\bf B 326} (1989) 108. 
\bibitem{Maldacena} J. Maldacena, 
{\em The large N limit of superconformal
field theories and supergravity}, 
hep-th/9711200.
\bibitem{GKP} S.S. Gubser, I.R. Klebanov and A.M. Polyakov, 
{\em Gauge theory correlators from noncritical string theory}, 
hep-th/9802109.
\bibitem{W2} E. Witten, {\em Anti-de Sitter space and holography},
hep-th/9802150; {\em Anti-de Sitter space, thermal phase transition
and confinement in gauge theories}, 
hep-th/9803131.
\bibitem{MV} C. Vafa, {\em Puzzles at large N}, 
hep-th/9804172.
\bibitem{CHvD} O. Coussaert, M. Henneaux and P. van Driel, {\em
Class. Quant. Grav.} {\bf 12} (1995) 2961.
%\bibitem{MM} M. Ba\~nados, T. Brotz and M. Ortiz, 
%{\em Boundary dynamics and the statistical mechanics of
%the 2+1 dimensional black hole},
%hep-th/9802076.
\bibitem{CH} O. Coussaert and M. Henneaux,{\em Phys. Rev. Lett.} {\bf
72} (1994) 183.
\bibitem{HS} M. Henningson and K. Sfetsos, {\em Spinors and the
adS/CFT correspondence}, 
hep-th/9803251.  
See also R. Leigh and M.
Rozali, {\em the large N limit of the $(2,0)$ superconformal field
theory}, 
hep-th/9803068 as well as the earlier work P. Breitenlohner
and D.Z. Freedman, {\em Ann. Phys. (N.Y.)} {\bf 144} (1982) 197.
\bibitem{Achucarro-Townsend} A. Ach\'ucarro and
P.K. Townsend, {\em Phys. Lett.} {\bf B180}, 89 (1986).
\bibitem{con} Our conventions for the spinors are the following. The
spinors
 are real so that $\bar \psi = \psi^t \gamma_0$. The Dirac matrices
 are
 taken to be
%\begin{equation}
$$
\gamma_0 = \left(\begin{array}{cc}   0 &  1 \\
                                    -1 &  0
                   \end{array} \right), \ \ \
\gamma_1 = \left(\begin{array}{cc}  1 & 0 \\
                                    0 & -1
                   \end{array} \right), \ \ \
\gamma_2 = \left(\begin{array}{cc}  0 & 1 \\
                                           1 &  0
                   \end{array} \right)
%                  \label{g0}
$$
%\label{Dm}
%\end{equation}
and satisfy
%\begin{equation}
$
~[ \gamma_a,  \gamma_b] = 2\epsilon_{abc} \gamma^c, %\ \ \ 
% \mbox{Tr}(\gamma_a\gamma_b) = 2\eta_{ab}, \ \ \ \
%(\gamma_0 \gamma_a)^t = \gamma_0 \gamma_a
%\end{equation}
$
where we use $\epsilon^{012}=1$. 
The covariant derivative $D$ in (\ref{a2}) acting on a spinor
$\lambda$  is
defined as $D \lambda = d\lambda + \frac{1}{2}A^a \gamma_a \lambda $
and
satisfies $D\wedge D \lambda = \frac{1}{2} F^a \gamma_a \lambda$ with
$F^a= dA^a + (1/2) \epsilon^a_{\ bc} A^b \wedge A^c$.
The generators of the super-algebra $OSp(1|2)$ are $3 \times 3$
supermatrices.  The bosonic generators are obtained by
augmenting the previous $J_a=(1/2)\gamma_a$ with one row and one
column of
zeros (they will still be denoted by $J_a$).  The fermionic
generators are
$$
e =  \left(\begin{array}{ccc}   0 &  0 & 1\\
                            0 &  0 & 0 \\
                            0 & 1 & 0
  \end{array} \right), \ \ \
f =  \left(\begin{array}{ccc}   0 &  0 & 0\\
                            0 &  0 & 1 \\
                        -1 & 0 & 0
\end{array} \right),
$$
and one has $e^2 = J_+, f^2 = - J_-$ where $J_\pm = J_2\pm J_0$.
Let $\psi$ be a $2$-component spinors and let $\Psi$ be the fermionic
supermatrix $\sqrt{2} \tilde{\psi} = \psi_1 e + \psi_2 f$.  Redefining
the product of anticommuting numbers by inserting an i, $ab \equiv i a
\cdot b$ -- so that $(a \cdot b)^{*} = a^{*} \cdot b^{*}$ --, 
one gets $sTr(\Psi \cdot \Xi) = i \bar{\psi} \xi$ and one
may rewrite the action (\ref{a2}) in the manifest super-Chern-Simons
form $(k/4 \pi) \int sTr( \Gamma \cdot d \Gamma + \frac{2}{3}
\Gamma\cdot \Gamma \cdot \Gamma)$ with $\Gamma = A + \Psi$, where
$sTr$ is the supertrace.  The supercurvature is ${\cal F} = d \Gamma +
\Gamma\cdot \Gamma$, and the equations of motion are just ${\cal F} =
0$.  The gauge transformations are $\delta \Gamma = d \Lambda + \Gamma
\cdot \Lambda - \Lambda \cdot \Gamma$ with $\Lambda \in osp(1|2)$.
\bibitem{RT} T. Regge and C. Teitelboim, {\em Ann. Phys. (N.Y.)}  {\bf
    88} (1974) 286.  
\bibitem{Max} M. Ba\~nados, {\em Phys. Rev.} {\bf
    D52} (1996) 5816 
\bibitem{MM} M. Ba\~nados, T. Brotz and M. Ortiz,
  {\em Boundary dynamics and the statistical mechanics of the 2+1
    dimensional black hole}, hep-th/9802076.  
\bibitem{HT} M. Henneaux
  and C. Teitelboim, {\em Commun. Math. Phys.}  {\bf 98} (1985) 391.
\bibitem{BTZ} M. Ba\~nados, C. Teitelboim and J. Zanelli, {\em Phys.
    Rev. Lett.} {\bf 69} (1992) 1849; M. Ba\~nados, M. Henneaux, C.
  Teitelboim and J. Zanelli, {\em Phys. Rev.} {\bf D 48} (1993) 1506.
\bibitem{Benguria} R. Benguria, P. Cordero, C. Teitelboim {\em Nucl.
    Phys.}{\bf B122} (1977) 61.  
\bibitem{Brown-Henneaux2} J.D. Brown
  and M. Henneaux, {\em J. Math. Phys.} {\bf 27} (1986) 489-491.
\bibitem{HT2} M. Henneaux and C. Teitelboim, ``Quantization
of Gauge Systems", Princeton University Press (Princeton: 1992),
chapter 2.
\bibitem{Balachandran} A.~P.\ Balachandran, L.~Chandar, and A.~Momen,
{\em Nucl.\ Phys.} {\bf B461} (1996) 581.
\bibitem{P5} A.M. Polyakov,
  {\em Int. J. Mod. Phys.} {\bf 5} (1990) 833.
\bibitem{P1} A. M. Polyakov, {\em Mod. Phys. Lett.} {\bf A2} (1987)
  893; V. G. Knizhnik, A. M. Polyakov and A. B.  Zamolodchikov, {\em
    Mod. Phys. Lett.} {\bf A3} (1988) 819; A. M. Polyakov and A.B.
  Zamolodchikov, {\em Mod. Phys. Lett.} {\bf A3} (1988) 1213.
\bibitem{Al} A. Alekseev and S. Shatashvili, {\em Nucl. Phys.}  {\bf B
    323} (1989) 719.  
\bibitem{Ber} M. Bershadsky and H. Ooguri, {\em
    Commun. Math. Phys.}  {\bf 126} (1989) 49.  
\bibitem{oR} P.
Forg\'acs, A. Wipf, J. Balog, L. Feh\'er and L. O'Raifeartaigh, {\em
      Phys. Lett.} {\bf 227 B} (1989) 214.
\bibitem{BersOo2} M. Bershadsky and H. Ooguri, {\em
Phys. Lett.} {\bf B 229} (1989) 374.
\bibitem{susy} M.T.
  Grisaru and R. M. Xu, {\em Phys. lett.} {\bf 205 B} (1988) 486; W.A.
  Sabra, {\em Mod. Phys. Lett.}  {\bf A 6} (1991) 875; {\em Nucl.
    Phys.} {\bf B 375} (1992) 82.  
\bibitem{Inami} T. Inami and K.I.
  Izawa, {\em Phys. Lett.}  {\bf B 255} (1991) 521.
\bibitem{Kni} V. G. Knizhnik, {\em Theor. Math. Phys.}
{\bf 66} (1986) 68.
\bibitem{Bersh2}M. Bershadsky, {\em Phys. lett.} {\bf 174} (1986) 285.
\bibitem{Fradkin} E. S. Fradkin and V. Ya. Linetsky, {\em Phys. lett.}
{\bf 291} (1992) 71.
\bibitem{Adem} M. Ademollo et al, {\em Phys. Lett.} {\bf
B 62} (1976) 105; {\em Nucl. Phys.} {\bf B 111} (1976) 77;
{\bf B 114} (1976) 297.
\bibitem{Ito} K. Ito, J. O. Madsen and J. L. Petersen,
{\em Extended Superconformal Algebras From Classical
and Quantum Hamiltonian Reduction}, hep-th/9211019.
\bibitem{Sevrin} A. Sevrin, K. Thielemans and W. Troost,
{\em Nucl. Phys.} {\bf B 407} (1993) 459.
\bibitem{Strominger} A. Strominger, {\em Black hole entropy from
    near-horizon microstates}, hep-th/9712251.  
\bibitem{Birmingham}
  D. Birmingham, I. Sachs and A. Sen, {\em Entropy of
    three-dimensional black holes in string theory}, hep-th/9801019.
\bibitem{Carlip} S. Carlip, {\em
    Phys. Rev.} {\bf D 51} (1995) 632. 
\bibitem{Bala} V. Balasubramanian and F. Larsen, {\em Near horizon
    geometry and black holes in four dimensions}, hep-th/9802198.
\bibitem{Kal} N.Kaloper, {\em Entropy count for extremal
    three-dimensional black strings}, hep-th/9804062.  
\bibitem{MSt}
  J. Maldacena and A. Strominger, {\em $AdS_3$ black holes and a
    stringy exclusion principle}, hep-th/9804085.  
\bibitem{Mar} E. Martinec, {\em Matrix models of
AdS gravity}, hep-th/9804111.
\bibitem{IPZ} M.
  Iofa and L. P. Pando Zayas, {\em Statistical entropy of Calabi-Yau
    black holes}, hep-th/9804129.  
\bibitem{BBG} K. Behrndt, I.
  Brunner and I. Gaida, {\em Entropy and conformal field theories of
    $AdS_3$ models}, hep-th/9804159.  
\bibitem{Cvetic} M. Cveti\v{c} and F. Larsen, {\em Microstates of
four-dimensional
rotating black holes from near-horizon geometry}, hep-th/9805146;
{\em Near-horizon geometry of rotating black holes in five-dimensions},
hep-th/9805097.
\bibitem{U1}
  S. Hyun, {\em U-duality between three and higher dimensional black
    holes}, hep-th/9704005 
\bibitem{U2} K. Sfetsos and K. Skenderis,
  {\em Microscopic derivation of the Bekenstein-Hawking entropy
    formula for non-extremal black holes}, hep-th/9711138; H. J.
  Boonstra, B. Peeters and K. Skenderis, {\em Branes intersections,
    anti-de Sitter spacetimes and dual superconformal theories},
  hep-th/9803231.  

\end{references}
\end{document}